\documentclass{aa}  

\bibpunct{(}{)}{;}{a}{}{,} 

\usepackage{graphicx}
\usepackage{txfonts}
\usepackage{natbib,twoopt}
\usepackage[breaklinks=true]{hyperref} 
\bibpunct{(}{)}{;}{a}{}{,}             
\makeatletter
  \newcommandtwoopt{\citeads}[3][][]{\href{http://adsabs.harvard.edu/abs/#3}%
    {\def\hyper@linkstart##1##2{}%
     \let\hyper@linkend\@empty\citealp[#1][#2]{#3}}}
  \newcommandtwoopt{\citepads}[3][][]{\href{http://adsabs.harvard.edu/abs/#3}%
    {\def\hyper@linkstart##1##2{}%
     \let\hyper@linkend\@empty\citep[#1][#2]{#3}}}
  \newcommandtwoopt{\citetads}[3][][]{\href{http://adsabs.harvard.edu/abs/#3}%
    {\def\hyper@linkstart##1##2{}%
     \let\hyper@linkend\@empty\citet[#1][#2]{#3}}}
  \newcommandtwoopt{\citeyearads}[3][][]%
    {\href{http://adsabs.harvard.edu/abs/#3}
    {\def\hyper@linkstart##1##2{}%
     \let\hyper@linkend\@empty\citeyear[#1][#2]{#3}}}
\makeatother

\begin{document}

   \title{Period, epoch and prediction errors of ephemeris from continuous sets of timing measurements}

   \author{H. J. Deeg
          \inst{}  }

   \institute{Instituto de Astrof\'\i sica de Canarias, C. Via Lactea S/N, E-38200 La Laguna, Tenerife, Spain\\
Universidad de La Laguna, Dept. de Astrof\'\i sica, E-38200 La Laguna, Tenerife, Spain\\
              \email{hdeeg@iac.es}
             }

   \date{Received 20 November 2014; accepted 14 March 2015}

 
  \abstract
  {Space missions such as \textit{Kepler} and \textit{CoRoT} have led to large numbers of eclipse or 
  transit measurements in nearly continuous time series. This paper shows how to obtain 
  the period error in such measurements from a basic linear least-squares fit, and how to 
  correctly derive the timing error in the prediction of future transit or eclipse 
  events. Assuming strict periodicity, a formula for the period error of such time series
   is derived: $\sigma_P = \sigma_T\  ({12}/{( N^3-N)})^{1/2}$, where  $\sigma_P$ is
    the period error; $\sigma_T$ the timing error of a single measurement and $N$ the
     number of measurements. Relative to the iterative method for period error estimation
      by Mighell \& Plavchan (2013), this much simpler formula leads to smaller period errors, 
      whose correctness has been verified through simulations. For the prediction of 
      times of future periodic events, the usual linear ephemeris where epoch errors are quoted 
      for the first time measurement, are prone to overestimation of the error of 
      that prediction. This may be avoided by a correction for the duration of the time series.
       An alternative is the derivation of ephemerides whose reference epoch and epoch 
       error are given for the centre of the time series. For long continuous or near-continuous 
       time series whose acquisition is completed, such central epochs should be the
        preferred way for the quotation of linear ephemerides. While this work was motivated from 
	the analysis of eclipse timing measures in space-based light curves, it should be 
	applicable to any other problem with an uninterrupted sequence of discrete timings
	 for which the determination of a zero point, of a constant period and of the
	  associated errors is needed.}


   \maketitle
%

\section{Motivation and objectives}
The space missions \textit{MOST}, \textit{Kepler} and \textit{CoRoT} have been dedicated to the acquisition 
of near-continuous photometry over longer time scales. From them, timing measurements 
of eclipse or transit events have become available of a different nature from those 
from ground-based campaigns. Their main difference is the completeness of coverage 
between the first and the last measurement, with duty cycles of about 90\% 
(\citeads{2013ASSP...31..145M}, for \textit{Kepler} and \textit{CoRoT\/}) over time scales ranging from weeks 
to years. The derivation of precise ephemerides for timing measurements of periodic 
events (typically eclipses or transits) in such data may therefore also require 
revised methods. In this paper, we  address two points:  the estimation
 of the period error on the one hand and the error in the prediction of the time of 
 future eclipse events - also denominated the `prediction error' - on the other. The derivation of the 
 prediction error is related to the correct usage of the epoch error of an ephemeris.  This 
 will lead also to a recommendation for an improved quoting of ephemeris from such 
 long-cadence time series.
 In all the work presented here, an intrinsically constant period of the observed 
 target is assumed.  An algorithmic method to derive period errors from continuous 
 series of timing measurements has been published by
  \citetads[][furthermore M\&P] {2013AJ....145..148M}. In the first part  
  (Sect.\ 2) of this communication, it is shown that a linear fit of the measured 
  individual transit or eclipse timings (or of 'O-C' residuals derived against 
  a preliminary ephemeris) is the correct way for a determination of a period and 
  its error, leading to a very simple equation to estimate the period error. The
   epoch error in a linear ephemeris is elaborated in the second part (Sect.\ 3). 
   This error, if quoted as usual for the first timing measurement in a dataset, is 
   shown to be a non-optimum description of the zero-point error in a linear ephemeris, 
   which may lead to overestimated prediction errors. Correct ways to estimate the 
   prediction error of future events are then given. 

\section{Derivation of the period error}
Our objective is the estimation of the error of the period $P$, given a continuous 
sequence of $N$ timing measurements $T_E$ at integer Epochs $E$, with $E=0,\dots,N-1$, 
and assuming that the period to be measured is intrinsically constant (e.g.\ $P$ does
 not vary with $E$). It is also assumed that all timing measurements have an identical 
 time error $\sigma_T$. A linear ephemeris given by 
\begin{equation}
T_{c,E} = P\cdot E + T_{c,0}\ , 
\label{eq:lineph}
\end{equation}
$T_{c,0}$ being the time of zero-epoch, can then be derived from minimizing 
the residuals $T_E - T_{c,E}$. We note that $T_E - T_{c,E}$ corresponds to the 
commonly used (O-C) or `observed - calculated' residuals.

The $\chi^2$ minimization to determine best-fit parameters for $P$ and $T_{c,0}$ is
 then given by the linear regression:
\begin{equation}
\chi^2 = \sum\limits_{E=0}^{N-1} \frac{(T_E - (P\ E + T_{c,0}))^2}{\sigma_T^2}.
\label{eq:chisq}
\end{equation}
The least squares estimate of the slope $b$ of a linear fit $y=a+bx$ to data-tuples 
$(x_i, y_i)$ can be found in many basic works on statistics (e.g.\ \citeads{Kenney1962, press+92}) and is given by:
 \begin{equation}
b=\frac{N \sum x_i y_i - \sum{x_i}\sum{y_i}}{N\sum{x_i^2}-(\sum{x_i})^2},
\label{eq:slope}
\end{equation}
where we use summations\footnote{Usually, summations over indices going from 1 to
 $N$ are assumed in Eq.~\ref{eq:slope} and also in Eq.~\ref{eq:intercept}. The 
 change to indices going from 0 to $N-1$ has no consequences as long as the 
 summations go over a total of $N$ terms. We prefer here indices starting with
  0 in order to start with E=0 in the linear ephemeris, as in Equation~\ref{eq:lineph}.} 
  over  $i = 0,\dots,N-1$. Recognizing that the values $x_i$ are given by the
   Epoch number $E$, and changing to the nomenclature of Eq.~\ref{eq:chisq} 
   ($a \rightarrow T_{c,0},\  b \rightarrow P, \ y_i \rightarrow T_{E}$) and 
   for the convenience of writing, replacing $E$ by $i$, we obtain:
 \begin{equation}
P=\frac{N \sum {i T_i - \sum{i}\sum{T_i}}}{N\sum{i^2}-(\sum{i})^2}.
\label{eq:fitP}
\end{equation}

It is of note that the $T_i$ in Eq.~\ref{eq:fitP} may  be either the measured times 
themselves (quoted for example in BJD) or  O-C' residuals against some other 
(preliminary) ephemeris. In that case, Eq.~\ref{eq:fitP} delivers the difference
 to the period of that ephemeris.
 
Making use of the identities $\sum_{i=0}^{N-1} i = N (N-1)/2$ and 
$\sum_{i=0}^{N-1} i^2 = N(N-1)(2N-1)/6\ $, we find, after some basic algebra:
  \begin{equation}
P=\frac{12 \ \sum{T_i\  (i - \frac{N-1}{2})}   }{N^3-N}={c_N}{\sum{T_i w_i}   }\ ; \  c_N={12}/  (N^3-N)
\label{eq:fitPsimp}
\end{equation}
The terms $w_i = i - \frac{N-1}{2}$ acts as weighting coefficients for 
the timings $T_i$, the highest weight being given for the timings at the beginning 
and end of a dataset, and with little or no weight for those near the centre.

From above equation for $P = P(T_0,...,T_{N-1})$ we can then derive the period 
error $\sigma_P$ using error propagation; e.g.\
 $\sigma_P^2 = \sum_i {(\sigma_T \frac{\partial P}{\partial T_i})^2} $, which leads immediately to:
  \begin{equation}
\sigma_P^2 = {c_N^2}{\sigma_T^2}\sum w_i^2 .
\label{eq:errprop}
\end{equation}
Using again the identities for $\sum_{i=0}^{N-1} i$ and $\sum_{i=0}^{N-1} i^2$, 
we find that $\sum w_i^2 = 1/c_N$ and arrive at the final result:
\begin{equation}
\sigma_P^2 =   {c_N}\ {\sigma_T^2} 
 = \frac{12\  \sigma_T^2}{ N^3-N}  \; \;; N\ge2 .
 \label{eq:Perr}
\end{equation}


\subsection{Comparison with the period errors of Mighell \& Plavchan}
In the following, period error estimates from Eq.~\ref{eq:Perr} are compared
 to similar ones given by M\&P. They use an iterative algorithm, denominated 
 `Period Error Calculator (PEC)' that obtains the period error through multiple 
 combinations of the errors of the manifold 2-point measurements that are 
 present within a series of timing measures. For $N=2$ to 8 timing measurements, 
 M\&P quote explicit values\footnote{See  the `Reduced' values in the table 1 of 
  M\&P. They give them for $M=1,..,7$ period cycles, which correspond to $N=2$,..,8 
  timing measurements} that correspond to the $c_N$ coefficients of  
  Eqs~\ref{eq:fitPsimp} or \ref{eq:Perr} given above.
A comparison of these values is shown in Table 1. Their results agree with 
Eq.~\ref{eq:Perr} only for the case of $N=2$ or $N=3$. Up to $N=8$, differences remain 
small within $\approx$30\%. Without implementing their PEC algorithm, we can also 
compare with their example of a strictly periodic variable with $N=171$ timing measurements, 
each with an uncertainty of $\sigma_T = 0.0104$ days, for which they derive a 
period error of 23 microdays.\footnote{See  M\&P's Figure 1 and accompanying 
text, which is for 170 cycles, corresponding to 171 measurements.} From our 
Eq.~\ref{eq:Perr}, with $c_{171} = 2.40 \cdot 10^{-6}$, we derive, however, a 
period error of 16.1 microdays, which implies that their `reduced value' 
for $N=171$ is $(23 / 16.1)^2 = 2.04$ times larger than $c_{171}$. 

\begin{table}
\caption{Comparison between $c_N$ values of this work and of Mighell \& Plavchan}
\label{table:1} 
\centering
\begin{tabular}{c c c}
\hline\hline
 & This work & M\&P\\ 
$N$\tablefootmark{a} & $c_N$& $c_N$\tablefootmark{b}\\
\hline 
2	&2	&2\\
3	&0.5	&0.5\\
4	&0.2	&0.22222\\
5	&0.1	&0.11806\\
6	&0.05714	&0.07125\\
7	&0.03571	&0.04574\\
8	&0.02381	&0.03163\\
$\vdots$&$\vdots$&$\vdots$\\
171	&2.40$\times 10^{-6}$	&4.89$\times 10^{-6}$\\
\hline 
\end{tabular}
\tablefoot{
\tablefoottext{a}{$N$ corresponds to $M+1$ in M\&P. } \tablefoottext{b}{Corresponding 
to the `Reduced' square of the period error estimates from M\&P's table 1.} 
}
\end{table}

\subsection{Simulations of the period error and  application notes}
\begin{figure}
   \centering
      \includegraphics[width=9.0cm]{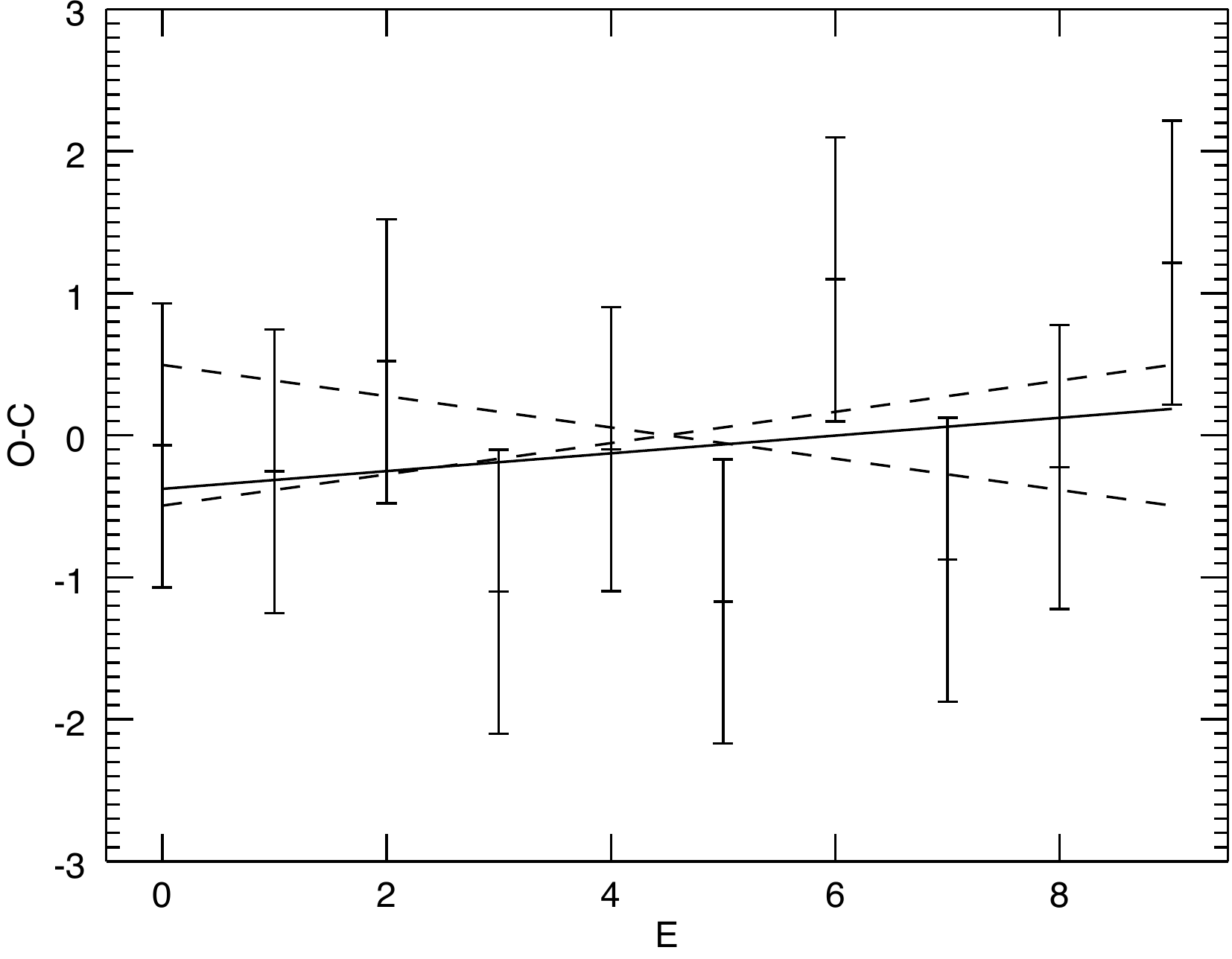}
   
   \caption{An `O-C'-style diagram of a simulation of $N=10$ timing measurements 
   drawn from a normal distribution with a standard deviation of $\sigma_T =1$ 
   against the intrinsic ephemeris, with individual errors  also of $\pm 1$. 
   The time unit is irrelevant in these simulations and any unit can be assumed. 
   The intrinsic (correct) ephemeris would be a horizontal line at $O-C=0$ 
   (not shown). The solid line shows the best fit to these measurements. The
    period error, given by Eq.~\ref{eq:Perr}, which is $\pm 0.1101$ time units per
     epoch, is indicated by the slopes of the dashed lines. In this particular simulation,{ the reduced chi-square against the intrinsic ephemeris was 0.64, }whereas the reduced chi-square against the fit was 0.75. }
              \label{fig:sim1}
    \end{figure}

\begin{figure}
   \centering
      \includegraphics[width=9.0cm]{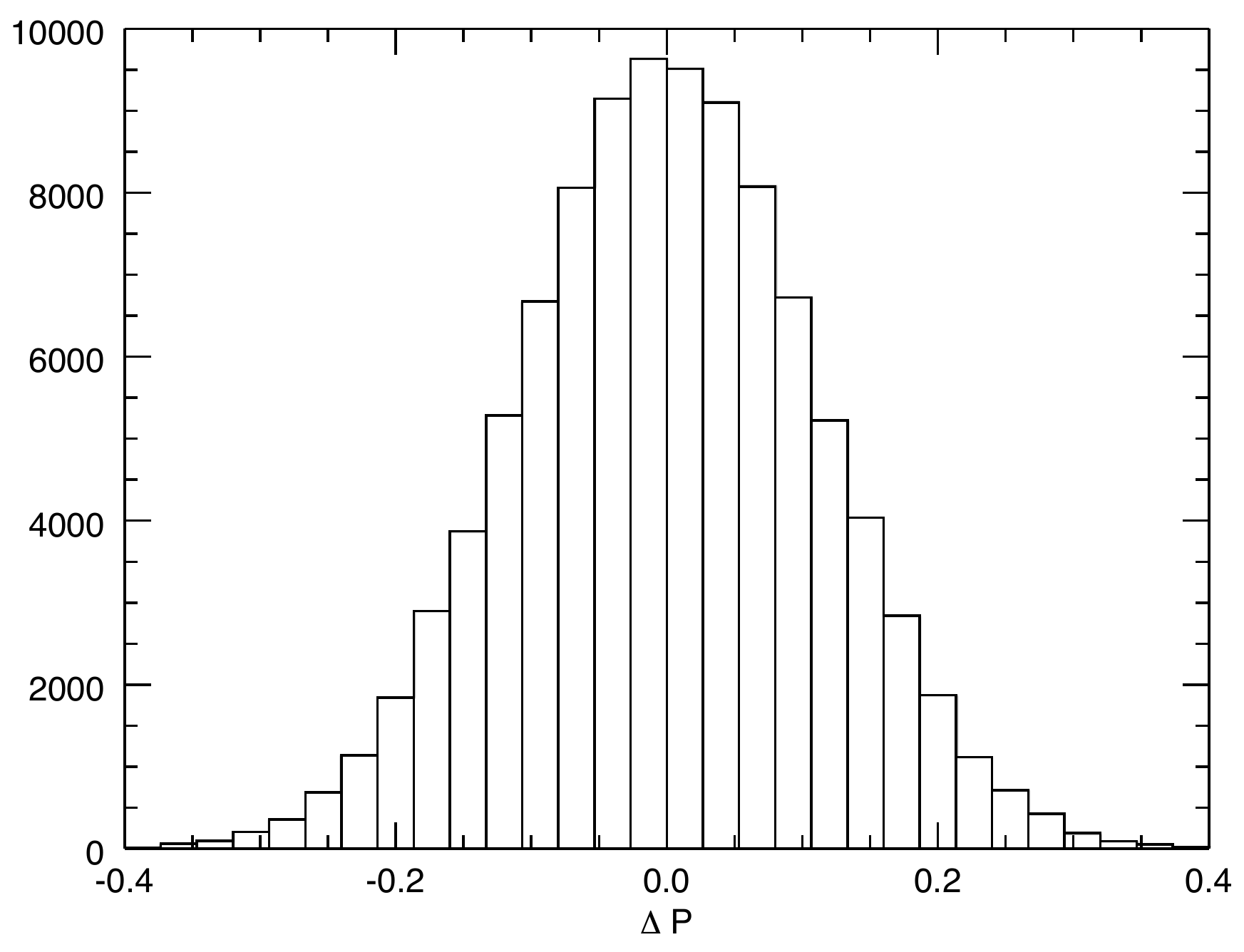}
\caption{Histogram of differences $\Delta P$ between fitted and intrinsic periods 
generated by repeating 100\ 000 times a simulation for $N=10$, as shown in 
Figure~\ref{fig:sim1}.  The standard deviation of $\Delta P$ is 0.1099, very close
 to the value from Eq.~\ref{eq:Perr} (0.1101 for $N=10$).}
              \label{fig:simhist}
    \end{figure}

Given that period error estimates from Eq.~\ref{eq:Perr} deviate significantly from 
those of the algorithm by M\&P, the results of Eq.~\ref{eq:Perr} were verified by 
a set of simulations as follows. Assuming that an intrinsic (correct) ephemeris 
is given by $T_c(E)=T_{c,0} + E\ P_c$, a set of $N$ timing measurements for $E=0,..,N-1$ 
is generated, with values $T_E$ that are randomly drawn from a normal distribution 
with a standard-deviation of  $\sigma_T$ and centred on the calculated 
value $T_c(E)$. The error of each individual timing measurement is also set to be $\sigma_T$.
 An example of such a simulated set of measurements in the form of an O-C diagram 
 against the intrinsic ephemeris is shown in Figure~\ref{fig:sim1}. A linear ephemeris 
 is then fitted by minimizing $\chi^2$ (e.g.\ as given by Eq.~\ref{eq:fitP} or  Eq.~ \ref{eq:fitPsimp}), which gives us the `observed period' $P_\mathrm{fit}$, 
  and the deviation against the intrinsic period, $\Delta P =  P_\mathrm{fit} - P_c$. 
  This procedure can easily be repeated many times using the same intrinsic ephemeris
  and a histogram of the deviations $\Delta P$ be produced, as shown in 
  Figure~\ref{fig:simhist}. The simulations show that, for large numbers of repetitions, 
  the mean of  $\Delta P$ becomes close to zero, and the standard deviation of 
  $\Delta P$ very closely approaches  the value predicted by Eq~\ref{eq:Perr}. 
  Such simulations with 100\ 000 repetitions were performed for values of 
  $N=3$, 10, 100, 1000, all of them showing the validity of the period error given by Equation~\ref{eq:Perr}.

\begin{figure}
   \centering
      \includegraphics[width=9.0cm]{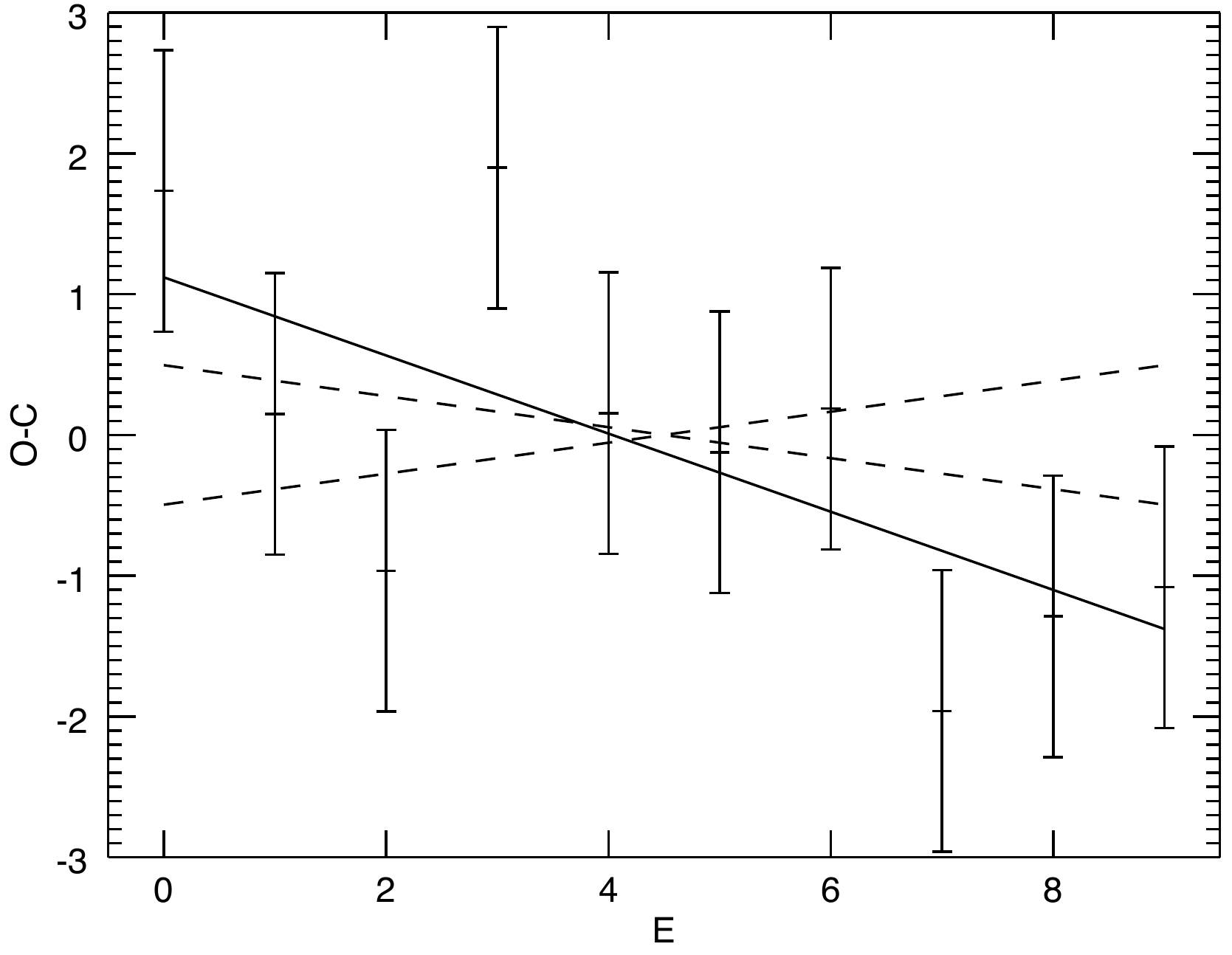}
   
   \caption{Like Fig.~\ref{fig:sim1}. In this particular simulation, a fortuitously good fit is obtained from data with a rather large scatter: The reduced chi-square against the intrinsic ephemeris was 1.43, whereas the reduced chi-square against the ephemeris fit (solid line) is 0.97. However, the fit's period has a rather large error, being 2.5 times as large as the period-error given by Eq.~\ref{eq:Perr} (dashed lines)}
              \label{fig:sim2}
    \end{figure}

Equation~\ref{eq:Perr} assumes that the measured times are normally distributed 
around the intrinsic (and unknown) linear ephemeris, and that the errors of 
the individual timing measurements (which are usually assigned by the observer) 
are of a size similar to the (O-C) residuals of the timings against this ephemeris. 
When this condition is not given, the reduced chi-square (using Eq.~\ref{eq:chisq} 
and $\chi^2_\mathrm{red} = \chi^2 /N $) against the \emph{intrinsic} ephemeris deviates 
significantly from $1$. {In practice, we know only the fitted ephemeris. The  reduced chi-square against it should be calculated as $\chi^2_\mathrm{red} = \chi^2 /(N-2) $, accounting for the two unknown fit parameters. If $\chi^2_\mathrm{red}$ deviates substantially from 1, there is in principle no way to know if this is due to unusually large or small random errors in the measurements or if it has other origins. For the case of $\chi^2_\mathrm{red}\ll  1$, we don't know if measurements have been fortuitously well aligned, or if measurement errors have been overstated. For $\chi^2_\mathrm{red} \gg 1$, measurement errors might have been understated.  However, an intrinsic period-variation could also be the origin for the poor fit --  a revision of the physical condition of the observed system should then be performed, to evaluate if that might be a valid hypothesis. Even if the fitted ephemeris indicates a good fit with $\chi^2_\mathrm{red} \approx 1 $, we have to be aware that a large $\chi^2_\mathrm{red}$ against the unknown true ephemeris may still be present, and that the fitted parameters may have errors larger than expected. This can happen due to measurements being fortuitously aligned to represent a rather deviant ephemeris. A simulation with such an outcome in shown in Fig.~\ref{fig:sim2} 
}


\section{Epoch and prediction errors}
In a similar way to that discussed for the period and period error  in the previous section, we 
can derive the intercept of the linear fit, which gives the ephemeris zero epoch $T_{c,0}$ and its error. 
\subsection{Epoch errors at the beginning of a measurement sequence}
First, we continue to use the usual epoch indices ranging from 0 to $N-1$ and start 
from a common equation$^1$ for the intercept:
 \begin{equation}
a=\frac{ \sum y_i \sum x_i^2 - \sum{x_i}\sum{x_i y_i}}{N\sum{x_i^2}-(\sum{x_i})^2}.
\label{eq:intercept}
\end{equation}
Changing, as before, to the nomenclature of Eq.~\ref{eq:chisq} and with very similar algebra, we obtain:
  \begin{equation}
T_{c,0}=\frac{6 \ \sum{T_i\  (\frac{2N-1}{3} - i)}   }{N^2+N}={d_N}{\sum{T_i v_i}   }\ ; \  d_N={6}/  (N^2+N).
\label{eq:fitT0}
\end{equation}
Again, terms $v_i = \frac{2N-1}{3}- i$ act as weighting coefficients for the timings 
$T_i$, with a weight that goes from about $\frac{2}{3} N$ through zero to $-\frac{1}{3}N$. For the 
error of $T_{c,0}$ we obtain similarly to Eq.~\ref{eq:errprop}: 
  \begin{equation}
\sigma_{T_{c,0}}^2=\sigma_P^2 = {d_N^2}{\sigma_T^2}\sum w_i^2 .
\end{equation}
Evaluating the sum of the weighting coefficients as \mbox{$\sum w_i^2 = N (N+1) (2N-1)/18 $}, we then obtain:
  \begin{equation}
\sigma_{T_{c,0}}^2 = \frac {\ (4N-2)\ \sigma_T^2}{N^2+N} .
\label{eq:errT0}
\end{equation}

\subsection{Ephemeris with zero epoch at the centre of the measurement sequence}
In the following, a zero epoch at the centre of the measurement sequence is 
considered. For simplicity, only the case with an odd number of  measurements  
is elaborated. The indices (epochs) of the measurements are now labelled $j$ 
and will be $j= -k,\dots,k$ with $k=\frac{1}{2}(N-1)$. For the period, we start 
again from the basic Eqs.~\ref{eq:slope} or rather \ref{eq:fitP}. For the summations, 
going now from $-k$ to $+k$, we employ the identities $\sum_{j=-k}^{k} j = 0 $ and 
$\sum_{j=-k}^{k}  k^2 = N(N+1)(N-1)/12\ $. The value for the period then turns  out as:
   \begin{equation}
P=\frac{12 \ \sum{T_j\  j }   }{N^3-N} .
\label{eq:fitPsimpz}
\end{equation}
Given that $j=i-\frac{N-1}{2}$, this is identical to Equation~\ref{eq:fitPsimp}. The 
equation for the period error is then of course also identical to Equation~\ref{eq:Perr}. 
For the intercept, or zero-epoch, and starting from Eq.~\ref{eq:intercept}, however,
we obtain a different and much simpler expression:
\begin{equation}
T_{c,m}=\frac{1}{N} \ \sum{T_i} .
\label{eq:fitTz}
\end{equation}
This `centre' or  `middle epoch' has been labelled  $T_{c,m}$ in order to 
distinguish it from the usual zero epoch, $T_{c,0}$, at the first measurement.
 The corresponding error is given by:
\begin{equation}
\sigma_{T_{c,m}}^2 = \frac {\sigma_T^2}{N} .
\label{eq:errTz}
\end{equation}
The different outcome for the epoch error, depending on its location at the beginning or 
in the centre of a measurement sequence, can be explained in the following way.
From the epoch error at the centre of a measurement sequence $\sigma_{T_{c,m}}$ 
and from the period error $\sigma_P$, we can can calculate the expected timing error 
at the beginning (or at the end) of the sequence, $\sigma_{T_{\textrm{begin}}}$, using 
the square-sum of errors:
\begin{equation}
\sigma_{T_\textrm{begin}}^2 = \sigma_{T_{c,m}}^2 + (\textstyle {\frac{N-1}{2} }\sigma_P)^2,
\label{eq:sigTbegin}
\end{equation}
where $\textstyle {\frac{N-1}{2}}$ is the number of periods between the sequence's 
beginning and its centre. Inserting the expressions for $\sigma_{T_{c,m}}$ and 
$\sigma_P$ into Eq.~\ref{eq:sigTbegin}, we find that $\sigma_{T_\textrm{begin}} = \sigma_{T_{c,0}}$, 
with $\sigma_{T_{c,0}}$ being given by Equation~\ref{eq:errT0}: This means that the error of the 
zero epoch of an ephemeris quoted in the conventional way, with $E=0$ corresponding to the first 
timed event, is really the error-sum of the `true' epoch error $\sigma_{T_{c,m}}$ in the
 middle of the sequence plus a contribution from the period error.

\subsection{Consequences for the prediction of events beyond the measurement sequence}
\begin{figure}
   \centering
      \includegraphics[width=9.0cm]{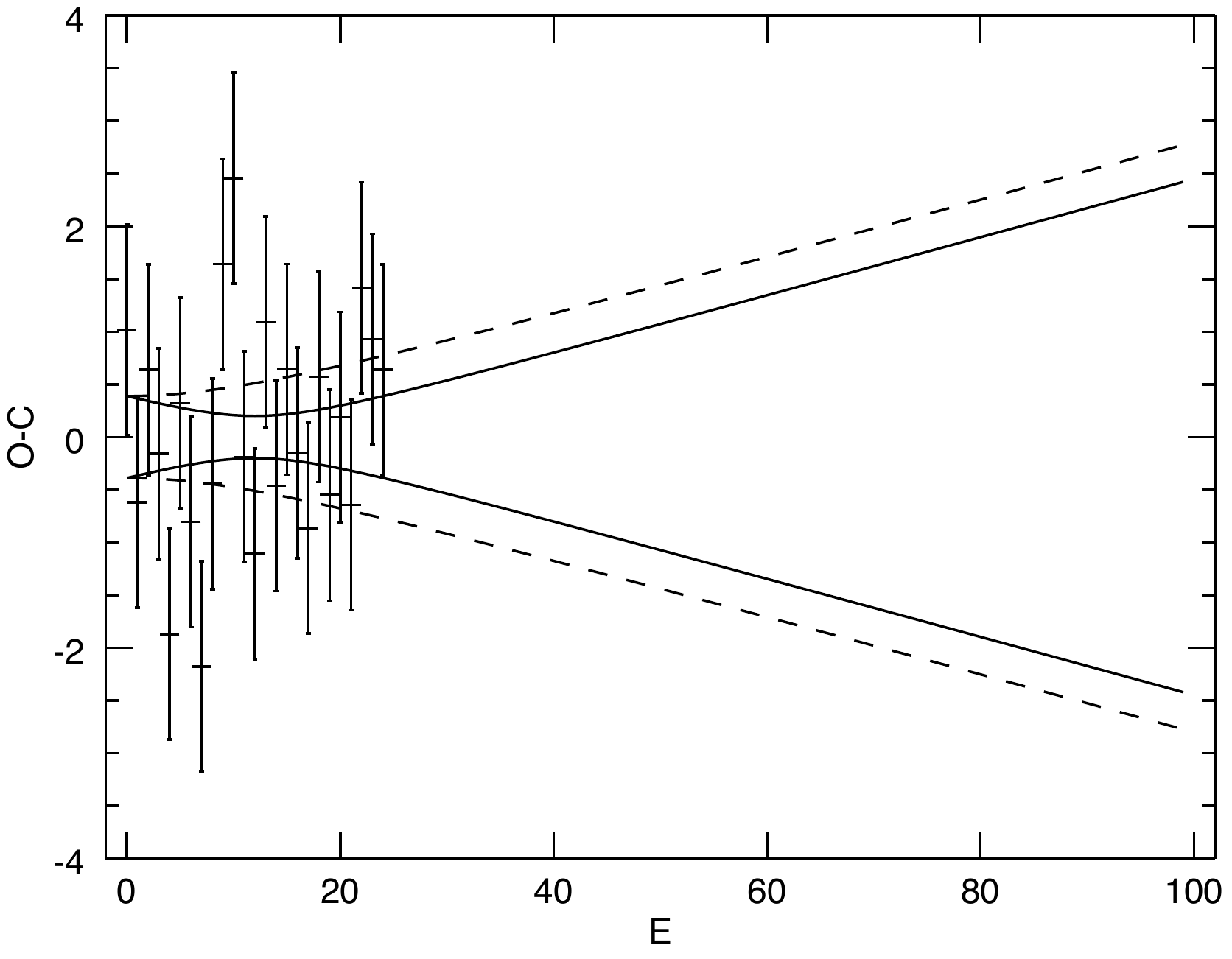}
   
   \caption{Expected timing uncertainties during and after a sequence of timing measurements 
   based on an ephemeris with given errors in epoch and period. The axes are similar to
    those in Figure~\ref{fig:sim1}. The dashed line gives the development of the 1-sigma uncertainty 
    from an error sum using the period error and the epoch error at the beginning of the 
    sequence. The solid lines outline the correct timing uncertainty, with the epoch 
    error being derived for the centre of the sequence. For epochs beyond the end of the 
    sequence, this uncertainty is also identical to the `prediction uncertainty' given 
    by Equation~\ref{eq:sigpred}.}       \label{fig:prederr}
    \end{figure}

 A fundamental function of an ephemeris is the prediction of transit or eclipse events 
 beyond the end of a measurement sequence, giving both the time and the time uncertainty 
 of such future events. The usually quoted $\sigma_{T_{c,0}}$  and $\sigma_P$ may, however, 
 easily lead to an overestimation of this timing error if a naive error-sum given by 
 $\sigma_{T_c,E}^2 = \sigma_{T_{c,0}}^2 + (E\ \sigma_P)^2$ is used, as is illustrated 
 by the dashed slopes in Figure~\ref{fig:prederr}.
There are two solutions to circumvent such overestimation of the `prediction error'.
The first solution, $\sigma_{T_{c,m}}$ can be retrieved\footnote{As a side-note to Eq.~\ref{eq:sigTzrev}, 
the calculation of $\sigma_{T_{c,m}}$ for existing ephemerides may also serve as a 
diagnostics on the correct sizing of these ephemeris errors since $\sigma_{T_{c,m}}$ 
can easily be related to the size of the errors of individual timing measurements 
through Eq.~\ref{eq:errTz}, and, of course, $\sigma_{T_{c,m}}^2$ needs to come out as a 
positive number --  or else $\sigma_{T_{c,0}}$ is underestimated and/or $\sigma_P$ is 
overestimated.} from the conventionally quoted epoch errors by reversing Eq.~\ref{eq:sigTbegin}:
\begin{equation}
\sigma_{T_{c,m}}^2 = \sigma_{T_{c,0}}^2 - (\textstyle {\frac{N-1}{2} }\sigma_P)^2
\label{eq:sigTzrev}
\end{equation}
The time-uncertainty at any epoch before, during, or after the measurement sequence 
can then be obtained from the error-sum:
\begin{equation}
\sigma_{T_{c,j}}^2=\sigma_{T_{c,m}}^2 + (j \ \sigma_P)^2 ,
\label{eq:sigTj}
\end{equation}
where $j$ is the number of epochs relative to the centre of the sequence.
A second, simpler, way to obtain a correct prediction error comes from the observation 
that the epoch error at the beginning or at the end of a measurement sequence should be 
identical. We can then use the conventional epoch error $ \sigma_{T_{c,0}}$ for the
 prediction of future events by counting from the end of the sequence, using:
\begin{equation}
\sigma_{T_{c,E}}^2 = \sigma_{T_{c,0}}^2 + (E - N)^2\sigma_P^2 \; ; E \ge N ,
\label{eq:sigpred}
\end{equation}
where $E$ is the epoch number since the begin of the sequence and $\sigma_{T_{c,E}}$ 
is the time uncertainty at $E$. This equation correctly takes into account the duration 
of the measurement sequence (see also Fig.~\ref{fig:prederr}). \\
 
\section{Conclusions}
In the first part of this communication, a simple formula for the derivation of 
period errors in continuous sequences of timing measurements with identical timing 
errors has been derived and verified through a set of simulations. For the extraction 
of a linear ephemeris from a set of timing measurements, which implies that an intrinsic 
constant period is assumed, there is no apparent reason to use other methods than a 
linear fit based on an error minimization. 
There is no reason to use another
 method for the estimation of the fit-parameters errors beyond an error propagation 
 from the equations that determine the fit parameters (in this case, from 
 Eqs~\ref{eq:slope} and \ref{eq:intercept}). In the case of identical measurement 
 errors in a continuous series of data, equations for these errors simplify to those 
 given in Sect.\ 2 for the period (Eq.~\ref{eq:Perr})  and Sect.\ 3 for the epoch 
 (Eqs~\ref{eq:errT0} and \ref{eq:errTz}). The only point open to variation is the
  type of error minimization used in the linear fit, where other methods beyond chi-square 
  minimization might be considered, such as minimizing absolute errors; and/or robust 
  fits that reject outliers. These may lead to slightly different error-estimates, all 
  of them, however, are based on the residuals against the best linear best fit, independently 
  on how that fit was obtained. 
  
It is not the aim of this communication to revise or analyse the `PEC' algorithm of M\&P, 
which derives the period error from a combination of timing errors between any pairing of 
two time measurements within a sequence. Certainly, the algorithm by M\&P is a vastly more 
complicated way to derive the period error. Differences between the error estimates from 
Eq.~\ref{eq:Perr} of this work and M\&P's `PEC' -- which increase with the number of 
timing measures -- {are probably caused by an incorrect weighting of the individual 2-point 
timing measurements from which PEC constructs its final result. } Verifying this would, however,
 need a detailed analysis of PEC, which is beyoned of the scope of this study. 

{
Both this work and the one by M\&P assume identical errors for all individual timing 
measurements. In practice, even in space missions such as \textit{CoRoT} and \textit{Kepler}, imperfect 
duty cycles cause occasional misses or incomplete transits. Furthermore, cosmic-ray hits
 may degrade light curves of individual transits, leading to larger timing errors. 
 For practical applications of the equations presented here, occasional measurements with
  strongly deviating timing errors (or missed measurements) can be ignored as long as a 
  predominant timing error can be identified. If such a predominant error cannot be
   identified (e.g.\ owing to a change of integration time in \textit{Kepler} light  curves), or if 
   a significant fraction of timing measurements is missing, the simplified equations 
   presented here will not be reliable. The ephemeris and its errors should then be 
   derived from a numerical least-squares minimization of Eq.~\ref{eq:chisq}, using 
   individual timing errors (e.g.\ \citeads{press+92}, sect.\ 14.2).}

In the second part of this article, two equations for the epoch error of continuous timing 
sequences are derived. In the first of these equations,the error is given for a `zero' epoch that corresponds 
to the first timing measurement (Eq.~\ref{eq:errT0}). This is the conventional way in 
which ephemerides are indicated. In the second equation, a much simpler expression is obtained for the 
epoch error at the centre of a timing sequence (Eq.~\ref{eq:errTz}). It is shown that 
these errors are equivalent,  the epoch error at the beginning (or end) of a timing 
sequence being the error sum of the central epoch error plus the period error. 
Ephemerides of long sequences of timing measurements would therefore be more logically 
expressed with epochs and epoch errors for a timing measurement at or near the sequence's 
centre. With such a `central ephemeris', the estimation of timing errors beyond the end 
of the original measurement sequence will then be correctly performed by a simple 
error sum between epoch error and the period error. The `central ephemeris' serves also 
to correctly estimate the uncertainties of the true (intrinsic) event times \emph{during} 
the measurement sequence. With conventional ephemerides, on the other hand, a correction for 
the duration of the measurement sequence is needed in order to derive correct timing 
errors for predictions past the measurement sequence. For the photometric follow-up of 
\textit{CoRoT} planets and planet candidates, Eq.~\ref{eq:sigpred} has been implemented for several
 years in an online calculator.\footnote{http://www.iac.es/proyecto/corot/followup/  
 (Access will be provided upon request to the author) } There, the numbers $N$ of
  observed transit events during the \textit{CoRoT} pointings are estimated from the target
   periods and the pointing durations, ranging from 28d to 159d. {Its predictions
    of future transit events have been shown to be reliable for \textit{CoRoT'}s planet 
    candidate verification programme (described initially in \citeads{2009A&A...506..343D}), 
    as well as for an ongoing  re-observation of \textit{CoRoT} planet transits (Klagyivik et al., 
    in prep.).} The use of Eq.~\ref{eq:sigpred} over naive error-sums of epoch and 
    period errors (counting the epochs since the beginning of the measurements) is still 
    more important when ephemerides of targets from the \textit{Kepler} mission are considered, 
    since in that case the difference between `naive' and correct prediction errors is 
    much larger, due to the 3.9 yr coverage of its light curves.

A successful re-observation of transits or eclipses of objects discovered by space 
missions such as \textit{CoRoT} or \textit{Kepler} depends critically on the correct prediction of their 
transit time and timing errors. Such predictions are essential  when only a few 
hours are available to observe a given transit or eclipse. A best possible derivation 
of the ephemeris errors and the prediction errors is therefore very important for the legacies 
of the space missions that have brought us these wonderful datasets of long, continuous 
and highly precise time series. {Precise ephemeris measurements may also be expected to
 have a similar impact on follow-up observations of  future planet detection missions, 
 namely \textit{TESS} \citepads{2014SPIE.9143E..20R} and \textit{PLATO} \citepads{2014ExA....38..249R}. }

\begin{acknowledgements}
  The author acknowledges support through grant AYA2012-39346-C02-02 of the Spanish Ministerio 
  de Econom\'ia y Competividad (MINECO). He also thanks the anonymous referee for his comments which led to an improvement in the presentation of this paper. 
     
\end{acknowledgements}


\bibliographystyle{aa} 
\bibliography{../../HJDmain,period_error7}

\begin{thebibliography}{7}
\expandafter\ifx\csname natexlab\endcsname\relax\def\natexlab#1{#1}\fi

\bibitem[{{Deeg} {et~al.}(2009){Deeg}, {Gillon}, {Shporer}, {Rouan},
  {Stecklum}, {Aigrain}, {Alapini}, {Almenara}, {Alonso}, {Barbieri}, {Bouchy},
  {Eisl{\"o}ffel}, {Erikson}, {Fridlund}, {Eigm{\"u}ller}, {Handler}, {Hatzes},
  {Kabath}, {Lendl}, {Mazeh}, {Moutou}, {Queloz}, {Rauer}, {Rabus}, {Tingley},
  \& {Titz}}]{2009A&A...506..343D}
{Deeg}, H.~J., {Gillon}, M., {Shporer}, A., {et~al.} 2009, \aap, 506, 343

\bibitem[{Kenney \& Keeping(1962)}]{Kenney1962}
Kenney, J.~F. \& Keeping, E.~S. 1962, in Mathematics of Statistics, Pt. 1, 3rd
  ed (Princeton, NJ: Van Nostrand), 252--285

\bibitem[{{Michel}(2013)}]{2013ASSP...31..145M}
{Michel}, E. 2013, in Advances in Solid State Physics, Vol.~31, Stellar
  Pulsations: Impact of New Instrumentation and New Insights, ed. J.~C.
  {Su{\'a}rez}, R.~{Garrido}, L.~A. {Balona}, \& J.~{Christensen-Dalsgaard},
  145

\bibitem[{{Mighell} \& {Plavchan}(2013)}]{2013AJ....145..148M}
{Mighell}, K.~J. \& {Plavchan}, P. 2013, \aj, 145, 148

\bibitem[{{Press} {et~al.}(1992){Press}, {Teukolsky}, {Vetterling}, \&
  {Flannery}}]{press+92}
{Press}, W.~H., {Teukolsky}, S.~A., {Vetterling}, W.~T., \& {Flannery}, B.~P.
  1992, {Numerical recipes in C. The art of scientific computing} (Cambridge:
  University Press, |c1992, 2nd ed.)

\bibitem[{{Rauer} {et~al.}(2014){Rauer}, {Catala}, {Aerts}, {Appourchaux},
  {Benz}, {Brandeker}, {Christensen-Dalsgaard}, {Deleuil}, {Gizon}, {Goupil},
  {G{\"u}del}, {Janot-Pacheco}, {Mas-Hesse}, {Pagano}, {Piotto}, {Pollacco},
  {Santos}, {Smith}, {Su{\'a}rez}, {Szab{\'o}}, {Udry}, {Adibekyan}, {Alibert},
  {Almenara}, {Amaro-Seoane}, {Eiff}, {Asplund}, {Antonello}, {Barnes},
  {Baudin}, {Belkacem}, {Bergemann}, {Bihain}, {Birch}, {Bonfils}, {Boisse},
  {Bonomo}, {Borsa}, {Brand{\~a}o}, {Brocato}, {Brun}, {Burleigh}, {Burston},
  {Cabrera}, {Cassisi}, {Chaplin}, {Charpinet}, {Chiappini}, {Church},
  {Csizmadia}, {Cunha}, {Damasso}, {Davies}, {Deeg}, {D{\'{\i}}az}, {Dreizler},
  {Dreyer}, {Eggenberger}, {Ehrenreich}, {Eigm{\"u}ller}, {Erikson}, {Farmer},
  {Feltzing}, {de Oliveira Fialho}, {Figueira}, {Forveille}, {Fridlund},
  {Garc{\'{\i}}a}, {Giommi}, {Giuffrida}, {Godolt}, {Gomes da Silva},
  {Granzer}, {Grenfell}, {Grotsch-Noels}, {G{\"u}nther}, {Haswell}, {Hatzes},
  {H{\'e}brard}, {Hekker}, {Helled}, {Heng}, {Jenkins}, {Johansen},
  {Khodachenko}, {Kislyakova}, {Kley}, {Kolb}, {Krivova}, {Kupka}, {Lammer},
  {Lanza}, {Lebreton}, {Magrin}, {Marcos-Arenal}, {Marrese}, {Marques},
  {Martins}, {Mathis}, {Mathur}, {Messina}, {Miglio}, {Montalban}, {Montalto},
  {Monteiro}, {Moradi}, {Moravveji}, {Mordasini}, {Morel}, {Mortier},
  {Nascimbeni}, {Nelson}, {Nielsen}, {Noack}, {Norton}, {Ofir}, {Oshagh},
  {Ouazzani}, {P{\'a}pics}, {Parro}, {Petit}, {Plez}, {Poretti}, {Quirrenbach},
  {Ragazzoni}, {Raimondo}, {Rainer}, {Reese}, {Redmer}, {Reffert},
  {Rojas-Ayala}, {Roxburgh}, {Salmon}, {Santerne}, {Schneider}, {Schou},
  {Schuh}, {Schunker}, {Silva-Valio}, {Silvotti}, {Skillen}, {Snellen}, {Sohl},
  {Sousa}, {Sozzetti}, {Stello}, {Strassmeier}, {{\v S}vanda}, {Szab{\'o}},
  {Tkachenko}, {Valencia}, {Van Grootel}, {Vauclair}, {Ventura}, {Wagner},
  {Walton}, {Weingrill}, {Werner}, {Wheatley}, \&
  {Zwintz}}]{2014ExA....38..249R}
{Rauer}, H., {Catala}, C., {Aerts}, C., {et~al.} 2014, Experimental Astronomy,
  38, 249

\bibitem[{{Ricker} {et~al.}(2014){Ricker}, {Winn}, {Vanderspek}, {Latham},
  {Bakos}, {Bean}, {Berta-Thompson}, {Brown}, {Buchhave}, {Butler}, {Butler},
  {Chaplin}, {Charbonneau}, {Christensen-Dalsgaard}, {Clampin}, {Deming},
  {Doty}, {De Lee}, {Dressing}, {Dunham}, {Endl}, {Fressin}, {Ge}, {Henning},
  {Holman}, {Howard}, {Ida}, {Jenkins}, {Jernigan}, {Johnson}, {Kaltenegger},
  {Kawai}, {Kjeldsen}, {Laughlin}, {Levine}, {Lin}, {Lissauer}, {MacQueen},
  {Marcy}, {McCullough}, {Morton}, {Narita}, {Paegert}, {Palle}, {Pepe},
  {Pepper}, {Quirrenbach}, {Rinehart}, {Sasselov}, {Sato}, {Seager},
  {Sozzetti}, {Stassun}, {Sullivan}, {Szentgyorgyi}, {Torres}, {Udry}, \&
  {Villasenor}}]{2014SPIE.9143E..20R}
{Ricker}, G.~R., {Winn}, J.~N., {Vanderspek}, R., {et~al.} 2014, in Society of
  Photo-Optical Instrumentation Engineers (SPIE) Conference Series, Vol. 9143,
  Society of Photo-Optical Instrumentation Engineers (SPIE) Conference Series,
  20

\end{thebibliography}

\end{document}